\documentclass[twocolumn,showpacs,preprintnumbers,aps,pra,amsmath,amssymb,floatfix]{revtex4-1}

\usepackage{graphicx}
\usepackage{dcolumn}
\usepackage{bm}
\usepackage{epsfig}
\usepackage{color}
\usepackage{natbib}
\usepackage{epstopdf}

\def\Xstate{$X^1\Sigma^+$}
\def\astate{$a^3\Sigma^+$}

\begin{document}

\title{Observation of Feshbach resonances between ultracold Na and Rb atoms}

\author{Fudong Wang}
\author{Dezhi Xiong}
\author{Xiaoke Li}
\author{Dajun Wang}
\email{djwang@phy.cuhk.edu.hk}
\affiliation{Department of Physics and Center for Quantum Coherence, The Chinese University of Hong Kong, Shatin, Hong Kong, China }

\author{Eberhard Tiemann}
\affiliation{Institut f\"{u}r Quantenoptik, Leibniz Universit\"{a}t Hannover, Welfengarten 1,  30167 Hannover, Germany}

\date{\today}

\begin{abstract}

We have successfully prepared an optically trapped ultracold mixture of $^{23}$Na and $^{87}$Rb atoms and studied their interspecies Feshbach resonances. Using two different spin combinations, several $s$-wave and $p$-wave resonances are identified by observing high inelastic losses and temperature increases for both species near resonant magnetic field values. The two $s$-wave resonances observed below 500 G between atoms in their lowest energy levels are possible candidates for Feshbach molecule association. Our results are well characterized by a coupled-channel model which also refines the ground state interaction potentials between $^{23}$Na and $^{87}$Rb. This work opens up the prospect for preparing ultracold ensembles of ground-state bosonic NaRb molecules which are chemically stable and can provide strong dipolar interactions.

\end{abstract}

\maketitle

There has been a tremendous effort in producing ultracold polar molecules for more than one decade \cite{EPJD04, Carr09}. The successful preparation of fermionic $^{40}$K$^{87}$Rb molecules near quantum degeneracy in 2008 is a landmark and also provides us so far the most promising scheme in this endeavor \cite{Ni08}. As both methods, atom cooling and Feshbach molecule association\cite{Chin10,Kohler06} are well developed techniques in ultracold quantum gas research, once proper transition frequencies for the stimulated Raman adiabatic passage (STIRAP) are found starting from Feshbach molecules, this scheme can reliably produce a variety of ground-state molecules \cite{Winkler07, Ospelkaus08, Ni08}. The first set of experiments with KRb molecules has provided a glimpse of the rich amount of physics which can be studied with ultracold samples of polar molecules \cite{Ospelkaus10,Ni10,Miranda11}. 

However, the majority of the recent theoretical proposals with dipolar interactions require a quantum degenerate sample \cite{Baranov08}. For KRb, the large inelastic loss induced by chemical reaction makes further cooling to degeneracy problematic. But it is believed that applying the same production scheme developed for KRb to other molecular species where the $AB + AB \rightarrow A_2 + B_2$ reaction is energetically forbidden at ultralow temperatures, Bose-Einstein condensates (BEC) or degenerate Fermi gases of polar molecules could be achieved. As it has been pointed out, alkali diatomic molecules NaK, NaRb, NaCs, KCs and RbCs satisfy this requirement \cite{Zuchowski10} and they are applied presently by different groups around the world \cite{Pilch09,Takekoshi12,Cho12,Wu12,Park12}.    

In this work we present our results on Feshbach resonances (FRs) between $^{23}$Na and $^{87}$Rb atoms. We choose to work with this combination because NaRb molecule has, besides its chemical stability, a rather large permanent electric dipole moment of 3.3 Debye \cite{Aymar05}. Another advantage is that several low lying electronic potentials of NaRb molecule, including the ground states \Xstate~and \astate, were already studied in detail \cite{Docenko04,Pashov05,Pashov06,Docenko07} and predictions of FRs could be made from these results \cite{Pashov05}. We have observed 3 $s$-wave resonances in two different spin combinations which are within 50 G of these original predictions. A full coupled-channel calculation identifies several other resonances being of $p$-wave nature. This work opens up the possibility of producing NaRb Feshbach molecules and is an important step toward our quest for ground-state polar molecules. 

Although Na and Rb are two of the most popular atoms for quantum gas research, their optically trapped ultracold mixture has never been produced to our knowledge. We prepare them for the first time in the same single-chamber setup as previously described for our Rb BEC production \cite{Xiong13}. Similar to that work, we use ultra-violet light induced atom desorption to enhance loading of the magneto-optical trap (MOT) for both Rb and Na. Because of the low Na saturated vapor pressure at room temperature, our Na MOT contains only several million atoms which is almost 50 times less than Rb. After optical pumping both Rb and Na atoms to their $\left|f = 1, m_f = -1\right\rangle$ hyperfine states, they are loaded into a magnetic quadrupole trap. Forced microwave evaporation is then performed on Rb, and Na is sympathetically cooled by Rb. A 1070 nm laser beam with 4 W power and a 65 $\mu$m beam waist is applied for the optical dipole trap (ODT). The focus of the ODT is displaced by 90 $\mu$m from the quadrupole trap center to partially suppress Majorana losses. In this way, we can achieve temperatures below 10 $\mu$K, albeit with low evaporation efficiency. The atoms are then loaded into a crossed ODT by superimposing a second laser beam also of 65 $\mu$m beam waist to the focus of the first trapping beam. The two ODT beams cross each other with an angle of 62$^{\circ}$ in the horizontal plane. A low homogeneous magnetic bias field is applied and both Rb and Na atoms can be transferred into their lowest hyperfine state $\left|1, 1\right\rangle$ simultaneously with a radio frequency (RF) adiabatic rapid passage. 

To further increase the mixture's phase-space density, we continue the evaporative cooling in the crossed ODT. At 1070 nm, the relative trap depths $U_{Rb}$/$U_{Na}$ is 3. When ramping down the laser powers, evaporation happens mainly for Na because it sees a much shallower trap. The heavier Rb is sympathetically cooled by Na. For the best result, it is thus beneficial to prepare initially more Na atoms than Rb in the crossed ODT. In the end, we typically have 1.5$\times$10$^5$ atoms for each species. The sample temperature is 1.5 (1.8) $\mu$K for Na (Rb). Atom numbers and temperatures are measured by standard absorption imaging method. The final trap frequencies are measured to be 2$\pi\times$(280, 310, 140) Hz and 2$\pi\times$(320, 355, 160) Hz, respectively. Correspondingly, the calculated peak density is 10$^{13}$/cm$^{3}$ (3$\times$10$^{12}$/cm$^{3}$) for Rb (Na).

\begin{figure}
\centering
\includegraphics[width=.95\linewidth]{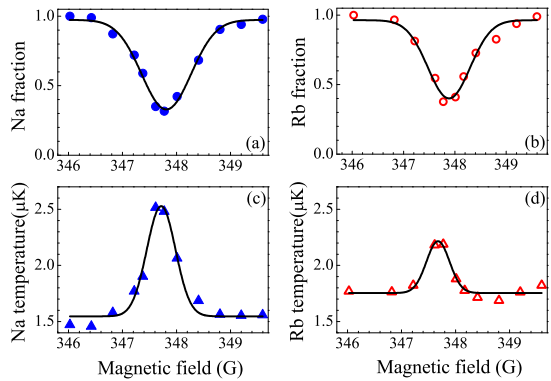}
 \caption{\protect\label{347G} (Color online) Feshbach resonance at 347.8 G for spin channel $f=1, m_f=1$ of both atoms, observed by loss and temperature increase on Na species, left side, and Rb species, right side.}
\end{figure}

Coarse searching of FRs is performed by first ramping the homogeneous magnetic field to a desired value and then applying a 2 Gauss sweep during the 500 ms holding time. The interspecies resonances manifest themselves as number losses and temperature increases in both species. We also confirm experimentally that atom losses and temperature changes do not occur for these field values with Na or Rb only. Once a resonance is located, the accurate resonance center is determined by the fast ramping and holding method with finer steps. The typical hold times are 50 ms in these final determinations. Observations were done over weeks and recordings were repeated to evaluate their reproducibility. We found an error of $\pm$0.25 G, deriving from long-term fluctuations in the setup. We calibrate the magnetic field by RF and microwave spectroscopy with a 0.1 G resolution on the precisely known Zeeman structure of Na or Rb and estimate the overall uncertainty to be $\pm$0.3 G taking reproducibility and calibration into account.

\begin{table*}
\caption{
Observed $^{23}$Na + $^{87}$Rb interspecies FRs below 1000 G for two different spin combinations with the assignments from coupled-channel calculations. The column ``I obs-cal'' gives the differences between observed and calculated resonance positions assuming two-body collisions applying fixed hyperfine interaction. The column ``II obs-cal'' gives the differences for $R$-dependent hyperfine interaction as modeled by eq. \eqref{eq:hfs}. The columns ``Type'' and ``Closed channel'' are explained in the text. Theoretical near resonance  scattering length profiles are calculated for $s$-waves at 1 nK and fitted to eq. \eqref{resonance} by the parameters $B_{res}$, $\Delta$ and $a_{bg}$. 
}
 \begin{center}
\begin{tabular}{c | c|c|c|c|c|c|c|c|c}
\hline \hline
\textrm{Entrance channel}& \textrm{B$_{exp}$} & $\Delta B_{exp}$ & I obs-cal& II obs-cal & Type & Closed channel&B$_{res}$& $\Delta$& a$_{bg}$(a$_0$)\\
 
\hline 
Na$|1,1>$ + Rb$|1,1>$  & 284.1(3) & 0.12(4) &-0.096& -0.051 & p M=1 and 2 $m_f$=2 & v=-2(2+1)&  &  &\\
   m$_f$=2 & 284.2(3)& 0.14(2) &-0.007& -0.024 & p M=3 $m_f$=2& &  &  &  \\
                & 284.9(3) & 0.04(1)  &-0.815& 0.112  & p M=2 $m_f$=1 & &  &  &\\
                & 285.1(3) & 0.06(1)  &-0.884& 0.068  & p M=1 $m_f$=1& &  &  & \\
                & 347.8(3) & 0.95(17) &-0.239& 0.009   & s M=2 $ f=3$& & 347.75 &-4.89  &66.77\\
                & 478.8(3) & 0.99(7)  &0.066&-0.039   & s M=2 $ f=2$& &478.79  &-3.80 &66.77\\
\hline
Na$|1,-1>$ + Rb$|1,-1>$ & 899.8(3)& 0.45(5) &-0.173& 0.022 & s M=-2 $f=2$ & v=-1(2+1)&899.82  & 0.333 &75.91\\
m$_f$=-2   & 954.2(3) & 0.12(1) &-0.342& -0.059 & p M=-2 $m_f$=-2 & &  &  &\\
               & 954.5(3) & 0.33(5) &-0.310&-0.026 & p  M=-3 and -1 $m_f$=-2& &  &  &\\
 
\hline \hline  
\end{tabular}
\label{table1}
\end{center}
\end{table*}

Two $s$-wave resonances at 347.8 G and 478.8 G were observed for Na $\left|1, 1\right\rangle$+Rb $\left|1, 1\right\rangle$ mixtures and Fig. \ref{347G} shows such recordings as an example for the loss and temperature variations. On resonance, more than 60$\%$ of the atoms are lost for both species and the sub-ensemble of each species also shows more than 30$\%$ temperature increase. These effects are attributed to the enhanced heteronuclear three-body collisions near an interspecies Feshbach resonance. We fit the atom loss profiles phenomenologically to a Gaussian function for extracting tthe resonance centers $B_{exp}$ and the full-width-half-maximum(FWHM) $\Delta B_{exp}$. The uncertainty from the profile fit is insignificant compared to the overall uncertainty. The $s$-wave resonances have measured widths of about 1 G. Another $s$-wave resonance is observed for Na $\left|1, -1\right\rangle$+Rb $\left|1, -1\right\rangle$ atoms at 899.8 G with a resonance width of only 0.4 G. As the derived widths are significantly influenced by three-body collisions, theoretical modeling is necessary to determine the two-body coupling strength between the open and closed channels.     

Based on the previously measured NaRb \Xstate~and \astate state potentials \cite{Pashov05}, a full coupled-channel model is constructed to fit the measured FRs. The Hamiltonian is set up as for several other heteronuclear alkali molecules\cite{Marzok2009}, and includes the Fermi contact interaction for Na and Rb and the atomic Zeeman effect. Initially, the atomic parameters for these interactions are directly incorporated as compiled by Arimondo et al \cite{Arimondo77}. With this first approach, we already can assign the FRs without ambiguity, as given in Table \ref{table1}. The column ``Type'' identifies the $s$- or $p$-wave character, the projection $M$ of the total angular momentum and the total angular moment $f$ of the coupled close channel to which the resonance correlates at zero field. The column ``Closed channel'' gives the vibrational assignment and the atomic asymptote (f$_{Na}$+f$_{Rb}$) from which the vibrational levels are counted, i.e. -1 represents the first level below the specified asymptote.

\begin{figure}
\centering
\includegraphics[width=.95\linewidth]{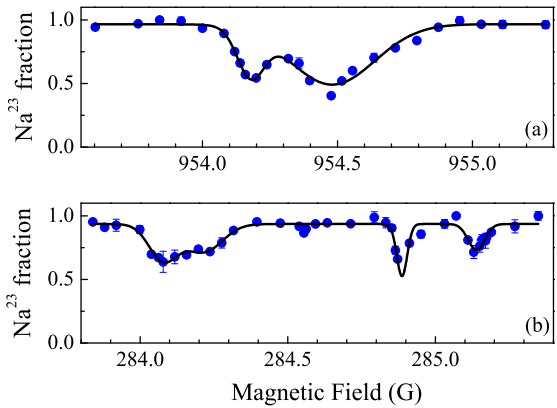}
 \caption{\protect\label{pwave} (Color online) $p$-wave FRs for two different entrance channels. (a) Na and Rb atoms are both in the $\left|1, -1\right\rangle$ hyperfine Zeeman level. (b) Na and Rb atoms are both in the $\left|1, 1\right\rangle$ hyperfine Zeeman level. Solid circles are measured remaining Na atoms at different applied magnetic fields. Few error bars shown reflect statistical spread of several shots at the same field. Solid lines are fits to multiple-peaks Gaussian functions for extracting the resonance centers. }
\end{figure}

The model also identifies several other observed features to be $p$-waves in nature. Detailed scans reveal their structures related to different projections m$_l$ of the rotational angular momentum $l$ onto the axis defined by the magnetic field and the resulting manifold of $M=m_f$ (prepared input channel) $ + m_l$. For the resonance around 954.5 G observed with the Na$\left|1, -1\right\rangle$+Rb $\left|1, -1\right\rangle$ spin combination, a 0.3 G splitting is clearly resolved as  shown in Fig. \ref{pwave} (a) by loss of Na atoms when the $B$ field is scanned near these resonances. Similar loss profiles for Rb atoms as well as heating of both species are also observed. In Fig. \ref{pwave} (b) $p$-wave resonances for the Na $\left|1, 1\right\rangle$+Rb $\left|1, 1\right\rangle$ entrance channel are shown. The left two features are centered at 284.1 and 284.2 G and are barely resolvable due to the limited magnetic field resolution and possibly to the 1.8 $\mu$K sample temperature. In Table \ref{table1} the type of resonance is specified by $m_f$ of the closed channel, resulting from the relation $M = m_f + m_l$. The manifold of possible $m_f$ is determined by the corresponding $f$ at zero field.  

For describing the $p$-wave splitting we have to include in the Hamiltonian the effective spin-spin interaction \cite{Knoop2011} resulting from the pure dipole-dipole coupling of the two electron spins and from second order spin-orbit interaction. We use a simple functional form for the effective molecular parameter $\lambda(R)$ describing the magnitude of this interaction as done e.g. in \cite{Knoop2011}:
\begin{equation}
\label{lambda} \lambda = -\frac{3}{4}
\alpha^2\left(\frac{1}{R^3}+a_\mathrm{SO}
\exp{\left(-bR\right)}\right).
\end{equation}  
Here $\alpha$ is the fine structure constant, $a_\mathrm{SO}$ gives the magnitude of the spin-orbit contribution and $b$ is set for the slope of the exponential to 0.7196 $a_0^{-1}$ which equals to the choice for Rb in \cite{Strauss2010} assuming that in NaRb the spin-orbit contribution will be determined by Rb. We fit the parameter $a_\mathrm {SO}$, which yields $a_\mathrm{SO}$ = -1.0(4) $a_0^{-3}$ from the observed splittings.  The contribution by second order spin-orbit coupling is small but significant at long range.

Precise determination of FRs relies on accurate potential curves for the \Xstate~and \astate~states of NaRb. Therefore, both potentials are constructed in a power series of the internuclear separation $R$, with the parametrization obtained from earlier spectroscopy work \cite{Pashov05}. In a series of iterations between coupled-channel calculations for the FRs and single channel calculations of rovibrational levels of the \Xstate~and \astate~states, the potentials are improved by fitting the underlying coefficients \cite{Tiemann2013} to match the locations of the FRs and simultaneously about 10000 rovibrational transitions from the Fourier-transform molecular spectroscopy \cite{Pashov05}. The calculated singlet and triplet levels responsible for the FRs are added as observables to the set of transitions with proper weights for joining the two fits.
  
In Table \ref{table1} we summarize all observed interspecies Na+Rb FRs together with their assignments from the coupled-channel calculations. The deviations between experiment and theory, shown in column ``I obs-cal'', are mostly within the experimental uncertainties. But the group of $p$-wave resonances around 285 G shows peculiarities within its five members (two resonances are almost degenerate, as given in the table). Looking into the detailed level scheme we find that this structure is a result of a strong mixing of the level $f=2$, $l=1$ and $v'=-1$ of the asymptote Na $\left|1\right\rangle$ + Rb $\left|1\right\rangle$ with $f=3$, $l=1$ and $v'=-2$ of the asymptote Na $\left|2\right\rangle$ + Rb $\left|1\right\rangle$. These levels are almost degenerate for $m_f=2$ at a field of 285 G. Thus the structure is very sensitive to the energy contributions by the potentials, the effective spin-spin coupling and the actual hyperfine interaction.  Assuming that the potentials and the effective spin-spin coupling are already quit well fixed, we studied the influence of a $R$-dependent hyperfine interaction. Such an effort is justified, because the two almost degenerate vibrational levels test such possible $R$-function differently. Such an effect has been extracted in other cases like Na \cite{Knoop2011} and Rb \cite{Strauss2010} through observable shifts of the FRs or bound level energies. 

The initially constant atomic parameter $a_A$ of the Fermi-contact interaction of atom $A$ is set up as a function of $R$:
\begin{equation}
\label{eq:hfs}
a_{A}(R)=a_{A,\rm hf}
         \left(1+c_f ~ e^{-R/dR}+ d_f\frac{(1-e^{R/dR})^{n+1}}{R^n}\right),
\end{equation}
which represents a changeover function  with width $dR$ to go from the atomic value $a_{A,\rm hf}$ at large $R$ to a molecular value $a_{M,\rm hf}(1+c_f)$ at small $R$. The additional exponent $(n+1)$ was chosen to avoid the divergence by the last term for $R \rightarrow$ 0. We use this changeover function for both atoms, Na and Rb. This type of function was found by several trials on the fit and $dR$ was chosen fairly large $dR=2.0 a_0$ and the exponent set to $n$ = 3. The parameter values $c_f$ and $d_f$ are determined in the fit to be $c_f$=-55.2 and $d_f$= 27.3$\cdot a_0^{n}$. These values give corrections on the hyperfine coupling in the order of few \% at long range. At short range the extrapolated magnitude becomes too large, but it is averaged out by the fast oscillations of the vibrational wavefunction for states close to the atom pair asymptote. Thus this approach only models the hyperfine interaction in the energy regime needed for describing the accidental degeneracy mentioned above and should not be applied to deeply bound states. 

The new fit is given by deviations in column ``II obs-cal'' of Table \ref{table1} and shows values smaller than the experimental uncertainty. We see that not only the modeling for the $p$-wave structures is improved, but also the $s$-wave resonances are better described. We should note that the effective number of variables for the FRs is now 6 compared to only 9 observations including the detailed structure of the $p$-wave resonances. This fact bears the risk of over fitting the observations. But the 3 $s$-wave resonances in connection with the positions of the $p$-wave groups determine already the potentials and the effective spin-spin contribution, comparing the results in column ``I obs-cal'' of the table. We believe that the overall improvement in the fit is a clear indication that the hyperfine interaction needs the extension in some form, e.g. the chosen one in eq. \eqref{eq:hfs}, but which is certainly not a unique choice. More experimental data are desirable to strengthen this point, e.g. precise measurements up to few MHz on deeply bound states like in the case of Rb$_2$ \cite{Strauss2010}. 

In the table we also give the resonance position $B_{res}$, the width $\Delta$ and the background scattering length $a_{bg}$ for characterizing the scattering length as a function of the $B$ field:
\begin{equation}
\label{resonance}
a=a_{bg}(1+\frac{\Delta}{B-B_{res}}+...),
\end{equation}
using as many terms as $s$-wave resonances are present in this entrance channel. $B_{res}$ (uncertainty $\pm$0.3G) of these profiles are not exactly the calculated ones in the fit because they were derived for a 1 nK temperature and the fit uses the peak of the elastic two-body rate. The derived $a_{bg}$ listed in Table \ref{table1} are valid in the vicinity of the resonances and should be accurate to 2\%. The scattering lengths of the uncoupled states become 106.74 $a_0$ for \Xstate~and 68.62 $a_0$  for \astate~with similar accuracy. 

In conclusion, we have observed and analyzed FRs between $^{23}$Na and $^{87}$Rb atoms. Two of the identified $s$-wave resonances with both atoms in their lowest energy levels provide favorable experimental conditions for producing Feshbach molecules with different closed channel characters (triplet(singlet) for the 347.8 G (478.8 G) resonance). These weakly bound molecules could be used in the future in our quest for ground-state NaRb molecules. Having Feshbach molecules with different triplet and singlet characters should enhance our chance of finding an excited-state level for an efficient STIRAP using either a pure singlet or a singlet/triplet mixed excited state to reach the lowest rovibrational level in the \Xstate state.

For accelerating future studies on NaRb we calculated the expected $s$-wave FRs of other channels for $f_{Na}$ = 1 and $f_{Rb}^{87}$ = 1 up to 1000 G:

$\left|1,0\right\rangle+\left|1,1\right\rangle$ : 484 G and 717 G

$\left|1,1\right\rangle+\left|1,0\right\rangle$ : 388 G and 519 G (i) and 796 G (i)

$\left|1,0\right\rangle+\left|1,0\right\rangle$ : 411 G and 526 G and 763 G (i)

$\left|1,-1\right\rangle+\left|1,1\right\rangle$ : 517 G and 727 G

$\left|1,1\right\rangle+\left|1,-1\right\rangle$ : 576 G (i) and 906 G (i)

\noindent The additional label (i) indicates resonances with significant inelastic two-body losses. The accuracy of these predictions is limited by the digital step size in the calculation to 1 G. Similar calculations can be made for other channels and even for Na$^{85}$Rb knowing the full potentials \cite{Pashov05}.

We acknowledge useful discussions with Li You. Our work is supported by Hong Kong RGC (CUHK403111, and Direct Grant 2060440).

\end{document}